\documentclass[ag]{copernicus}

\usepackage{color}

\begin{document}

\title{Downward auroral currents from the reconnection Hall-region}

\author[1,2]{R. A. Treumann\thanks{Visiting the International Space Science Institute, Bern, Switzerland}}
\author[3]{R. Nakamura}
\author[3]{W. Baumjohann}

\affil[1]{Department of Geophysics and Environmental Sciences, Munich University, Munich, Germany}
\affil[2]{Department of Physics and Astronomy, Dartmouth College, Hanover NH 03755, USA}
\affil[3]{Space Research Institute, Austrian Academy of Sciences, Graz, Austria}

\runningtitle{The downward auroral current generator}

\runningauthor{R. A. Treumann, R. Nakamura, and W. Baumjohann}

\correspondence{R. A.Treumann\\ (rudolf.treumann@geophysik.uni-muenchen.de)}

\received{ }
\revised{ }
\accepted{ }
\published{ }


\firstpage{1}

\maketitle

\begin{abstract}
We present a simple (stationary) mechanism capable of generating the auroral downward field-aligned electric field {that is} needed for {accelerating the ionospheric electron component up into the magnetosphere and confining the ionospheric ions at low latitudes (as is required by observation of an ionospheric cavity in the downward auroral current region). The lifted ionospheric electrons carry the downward auroral current. Our model is based on the assumption of collisionless reconnection in the tail current sheet. It makes use of the dynamical difference between electrons and ions in the ion inertial region surrounding the reconnection {\sf X}-line which causes Hall currents to flow. We show that the spatial confinement of the Hall magnetic field and flux to the ion inertial region centred on the {\sf X}-point  generates a spatially variable electromotive force which is positive near the outer inflow boundaries of the ion inertial region and negative in the central inflow region. Looked {at} from the ionosphere it functions like a localised meso-scale electric potential.} The positive electromotive force gives rise to upward electron flow from the ionosphere {during substorms (causing `black aurorae')}. A similar positive potential is identified on the earthward side of the fast reconnection outflow region which has the same effect,  explaining the observation that auroral upward currents are flanked from both sides by narrow downward currents.  

\end{abstract}

\subsection*{Introduction}
{There is no consensus yet on the origin of the substorm auroral field-aligned current system. Observations \emph{in situ} the Earth's auroral region have shown that it consists of a spatially separated combination of upward and downward field aligned currents which close in the ionosphere via Pedersen currents flowing perpendicular to the magnetic field at altitudes where the perpendicular ionospheric conductivity is large  \citep{elphic1998,carlson1998}. These currents are carried by electrons of different energies which have been accelerated along the magnetic field by field-aligned electric potential drops. }

{The origin of the downward current electrons is clearly ionospheric, implying that the (cold) electrons have been accelerated upward \citep{carlson1998,carlson1998a} along the magnetic field by the presence of an electric field that points down from the magnetosphere into the ionosphere\citep[see also the review in][]{paschmann2003}, being responsible for the absence of any optical aurora, i.e. causing `black aurora' \citep[for a review cf., e.g.,][who also provides a timely account of electron fluxes and field-aligned currents based on Freja and Fast observations in the downward current region]{marklund2009}. At the same time this strong magnetically parallel electric field keeps the ionospheric ion component in this region (of the auroral downward field-aligned current) down, confining it to low ionospheric altitudes thus causing a so-called ionospheric trough or plasma cavity.
Conversely the upward (or return) currents are carried by high energy (hot) downward flowing electrons which have been accelerated somewhere in the magnetosphere and are responsible for the known auroral phenomena.}

\begin{figure}[t!]
\centerline{{\includegraphics[width=0.5\textwidth,clip=]{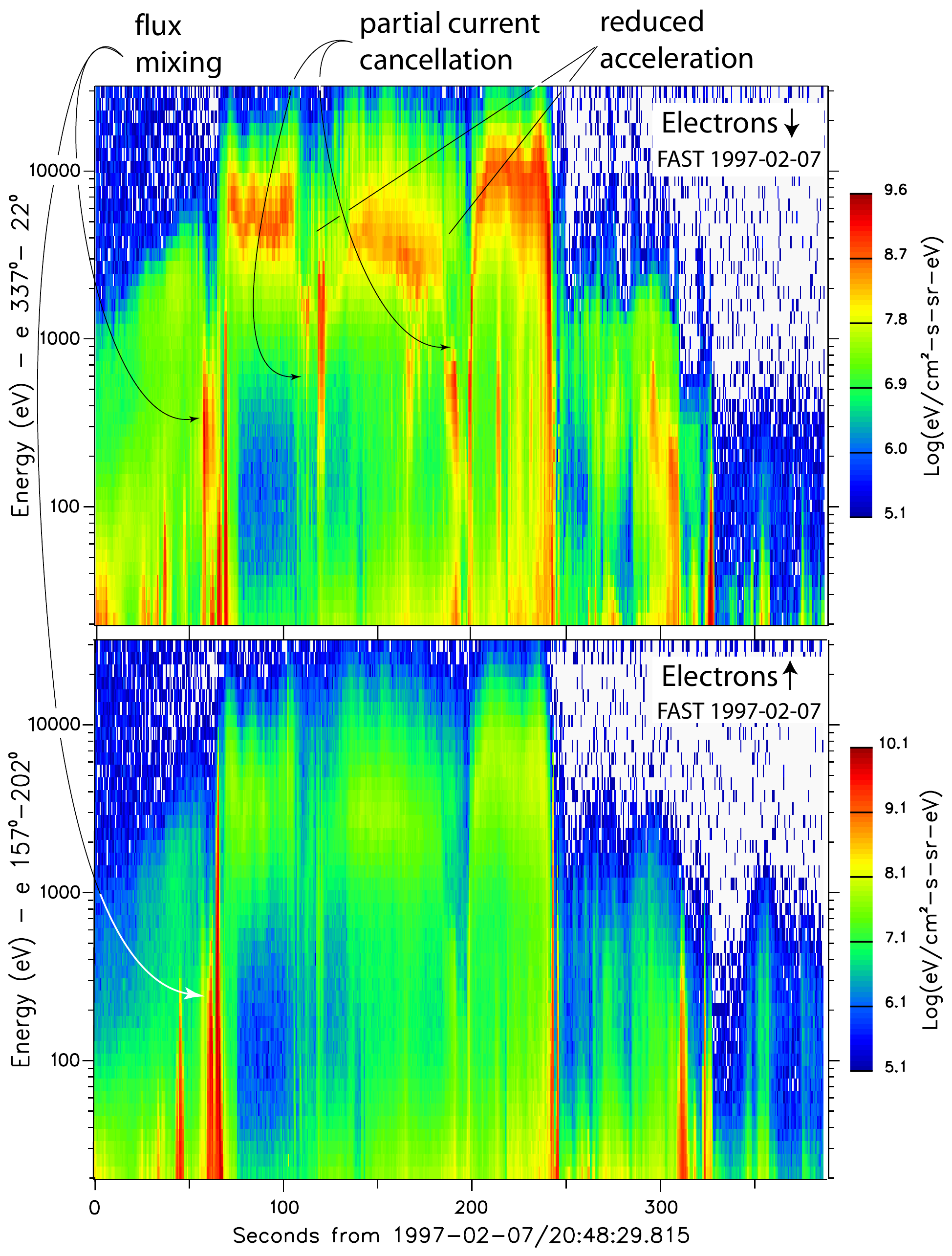}
}}
\caption[ ]
{\footnotesize {A representative irregular sequence of (low temporal resolution) auroral upward and downward electron fluxes (observed  by Fast in the northern hemispherical auroral region) during active substorm aurorae. The upper panel shows the measured electron fluxes within an angle of $\sim20^\circ$ centred on the direction parallel to the geomagnetic  field, the lower panel is centred on the direction anti-parallel to the geomagnetic field. Within the angular resolution of the instrument the asymmetry in the intensity of the data in both panels confirms the division of the auroral region into latitudinally adjacent regions dominated by either upward or downward electron fluxes corresponding to downward or upward field aligned currents, respectively. Note also the differences in electron energy. Upward electrons have low energies with maximum flux below 100 eV, i.e. accelerated from ionospheric electrons, while downward electrons dominate at several keV energy, i.e. being of magnetospheric origin \citep[figure taken from][]{treumann2009}.} }\label{aurorec-fig1a}
\vspace{-0.3cm}
\end{figure}

{These two regions of field-aligned currents and electron fluxes during aurorae are spatially separated. Usually one observes that the downward current-upward electrons form the latitudinally narrow northern (high-latitude) edge of the auroral disturbances while the upward current-downward electrons are found in a latitudinally much broader region adjacent to the former on its low-latitude side but separated from it by a narrow latitudinal gap where both field-aligned currents and electron fluxes are lacking. In addition, in most cases this latitudinally extended upward current-downward electron region is flanked on it southern side by another latitudinally narrow region of downward current-upward electron fluxes. This sequence seems to be observationally established though frequently obscured by the high dynamics of a substorm where many such regions can occur in a chain. Examples (cf. also Figure \ref{aurorec-fig1a}) can be found in the above cited literature.}

Recently we argued that the downward electron fluxes originate in the tailward reconnection region and that the complex structure of fluxes observed in the auroral region is caused by multiple reconnection sites in the magnetotail current sheet \citep[][]{treumann2009}. 
{In this Letter we investigate just one such reconnection site. We argue that reconnection in the geomagnetic tail may indeed generate both the fluxes of accelerated downward electrons which cause the active aurora and also may generate, which so far has neither been realised nor suspected, the field aligned electric fields that are needed if one wished to extract the ionospheric electrons from the ionosphere upward into the magnetosphere that carry the downward field-aligned auroral currents. For this to work, we find that reconnection must indeed be collisionless, occur in a sufficiently thin current sheet of width at most a few ion inertial lengths $\lambda_i$ (for the definition see below), and Hall currents must flow in the reconnection environment. These statements are not completely independent since Hall currents can flow in reconnection\footnote{The reader should note that we do not use the term Hall reconnection as this term is occupied by prticular versions of two-fluid MHD approaches which include the flow of Hall currents in one or the other way.} only under complete non-collisionality \citep{sonnerup1979,fujimoto1997,oieroset2001,nagai2001,treumann2006}.}
\begin{figure*}[t!]
\centerline{{\includegraphics[width=0.9\textwidth,clip=]{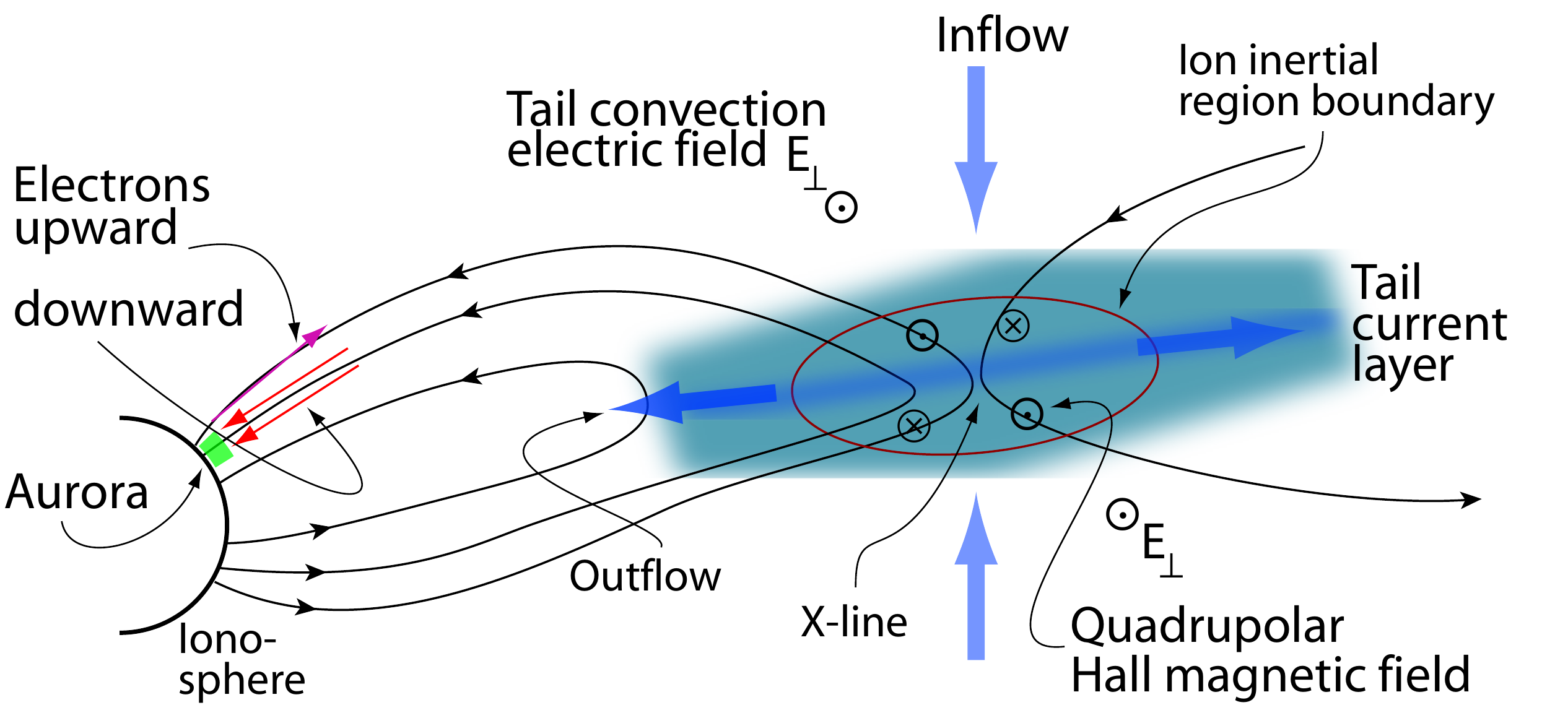}
}}
\caption[ ]
{\footnotesize {Schematic representation of the magnetic connection between the reconnection site in the geomagnetic tail current sheet, the ion-inertial region, convection electric field, Hall magnetic field, and the auroral region with the upward and downward electron fluxes (drawing not to scale!).} }\label{schematic-fig}
\vspace{-0.3cm}
\end{figure*}

\subsection*{Electromotive force in the reconnection-Hall region}
{Reconnection under collisionless conditions in thin current sheets of width the order of the ion-inertial scale $\lambda_i=c/\omega_{pi}$ has observationally been confirmed  \citep[see, e.g.,][]{fujimoto1997,nagai2001,oieroset2001,nakamura2006} to occur inside a region of Hall current flow.  These currents are carried by the magnetised plasma sheet electrons that flow across the non-magnetised plasma sheet ions \citep[for a pedagogically instructive sketch the reader may consult the paper by][]{treumann2006}. }

{The physics behind the generation mechanism of the Hall current under collisionless conditions is extraordinarily simple and may be summarised in a few sentences as follows: Hall currents require the presence of a magnetically perpendicular electric field ${\bf E}_\perp$ which forces the particles into an ${\bf E}_\perp\times{\bf B}$-drift perpendicular to both the magnetic and electric fields. This drift under non-collisionality is normally performed by both particle components and is current-free. (In the presence of collisions the ions are braked, only the electrons can flow, and a Hall current is caused, for instance in two-dimensional metals or in space in the ionosphere and probably as well in the solar chromosphere). In the collisionless state the mechanism of separation between electron and ion motions is due to the presence of a reconnection {\sf X}-point (or line) where the two plasma counter-streaming plasmas `collide'. Under these conditions classical Hall currents are secondary to reconnection, being just a side effect. Within a distance of one ion gyro-radius (or ion inertial length $\lambda_{ci}=c/\omega_{pi}$ with $\omega_{pi}^2=e^2N/\epsilon_0m_i$ the square of the ion plasma frequency) away to both sides of the {\sf X} line produced in reconnection, the ions behave non-magnetic. Transport of the magnetic field is the solely due to the ${\bf E}\times{\bf B}$ drift of the electrons against the ions. Hence, the electrons, in this region, carry a Hall current of strength $|{\bf J}_H|=e NE_\perp/B$. Due to their inertia, the ions are still slowly flowing in but do not perform an electric field drift anymore on this scale, rather they become accelerated along ${\bf E}_\perp$ thereby amplifying the sheet current (which may support the onset of reconnection).}

{In the magnetic {\sf X}-point geometry at the reconnection site these Hall currents generate a secondary quadrupolar Hall-magnetic field the geometry of which had been inferred thirty years ago by \citet{sonnerup1979} who also argued that the Hall current system necessarily included field-aligned currents in order to `close' it, connecting the Hall currents to the ionosphere. In fact, these field aligned closure currents \emph{do not belong} to the Hall current system; they only exist in the magnetospheric geometry where the field lines are tied to the conducting ionosphere. Hall currents are always exactly transverse to both the magnetic and the perpendicular electric fields. In an infinitely extended plane homogeneous collisionless current sheet field-aligned currents would be absent leaving reconnection independent even if including Hall currents. The divergence of the Hall currents in this case would be taken care of by their vanishing at the boundaries of the reconnection region. Reconnection in such plane current sheets has been studied extensively with the help of numerical PIC simulations \citep[e.g., by][and others]{zeiler2000,zeiler2002,scholer2003}. In recent papers \citep{baumjohann2010,treumann2010,treumann2011} we discussed the generation of seed-{\sf X} points and some aspects of its micro-scale physics. }

In the Earth's Magnetosphere-Ionosphere system connection between the Hall region and the ionosphere is, however, unavoidable {because of the presence of the conducting ionosphere with its high perpendicular conductivity. Figure \ref{schematic-fig} shows a very crude schematic of the magnetic connection between the tail reconnection site and the auroral region, i.e. the ion inertial domain at the reconnection site and the observed upward/downward electron fluxes in the auroral zone, both indicated in the figure. }

The downward auroral electron fluxes respectively upward (so called return) currents can in the reconnection model be understood as having their source in the acceleration mechanism acting at the reconnection site. No convincing reason could so far been given for the generation of the strong observed upward electron fluxes. Any near Earth models simply assume that a `battery effect' exists at the upper boundary of the ionosphere causing the required large field-aligned potential drops. This battery is assumed to be wave-driven, for instance by kinetic or shear Alfv\'en waves, or shear-flow driven, lacking any convincing reason for the appearance of shear flows at the top of the ionosphere.

{The conventional non-resistive model of how a field-aligned potential drop can be created under auroral conditions is based on the assumption that at the upper altitude boundary of the ionosphere some mechanism causes $\mathbf{E}_\perp\times\mathbf{B}$-shear flows \citep{carlson1998,carlson1998a,elphic1998}. Shear flows under collisionless conditions correspond to a spatial dependence of the perpendicular electric fields $\mathbf{E}_\perp(\mathbf{x})$. If these electric fields diverge in some place such that $\nabla\cdot\mathrm{E}<0$, their spatial dependence corresponds to a net `positive space charge' (potential) which attracts electrons and repulses positive ions along the magnetic field. Otherwise, if the electric field converges, the correspondence is to a net `negative space charge' which reflects electrons and attracts positive ions. In fact, shear flows of this case have barely been observed in the topside ionosphere. The interesting question that arises is, whether they may exist at the reconnection site in die near-Earth magnetosphere.We suggest here that the mere existence of the Hall (ion inertial) region at the near-Earth tail reconnection site is sufficient for producing the required `shear flow' electromotive forces, i.e. the field-aligned potential drop for extracting and upward accelerating the ionospheric electron component.}

{In order to attract electrons upward from the ionosphere as suggested by the model that has been decribed above (Figure \ref{schematic-fig}), one needs to generate the equivalent of a positive space charge at the lobe boundary of the Hall region.} In this section we demonstrate that the Hall region naturally produces such an induced equivalent space charge. Proof of this conjecture is quite easy to perform and proceeds along the following lines.

\begin{figure}[t!]
\centerline{{\includegraphics[width=0.4\textwidth,clip=]{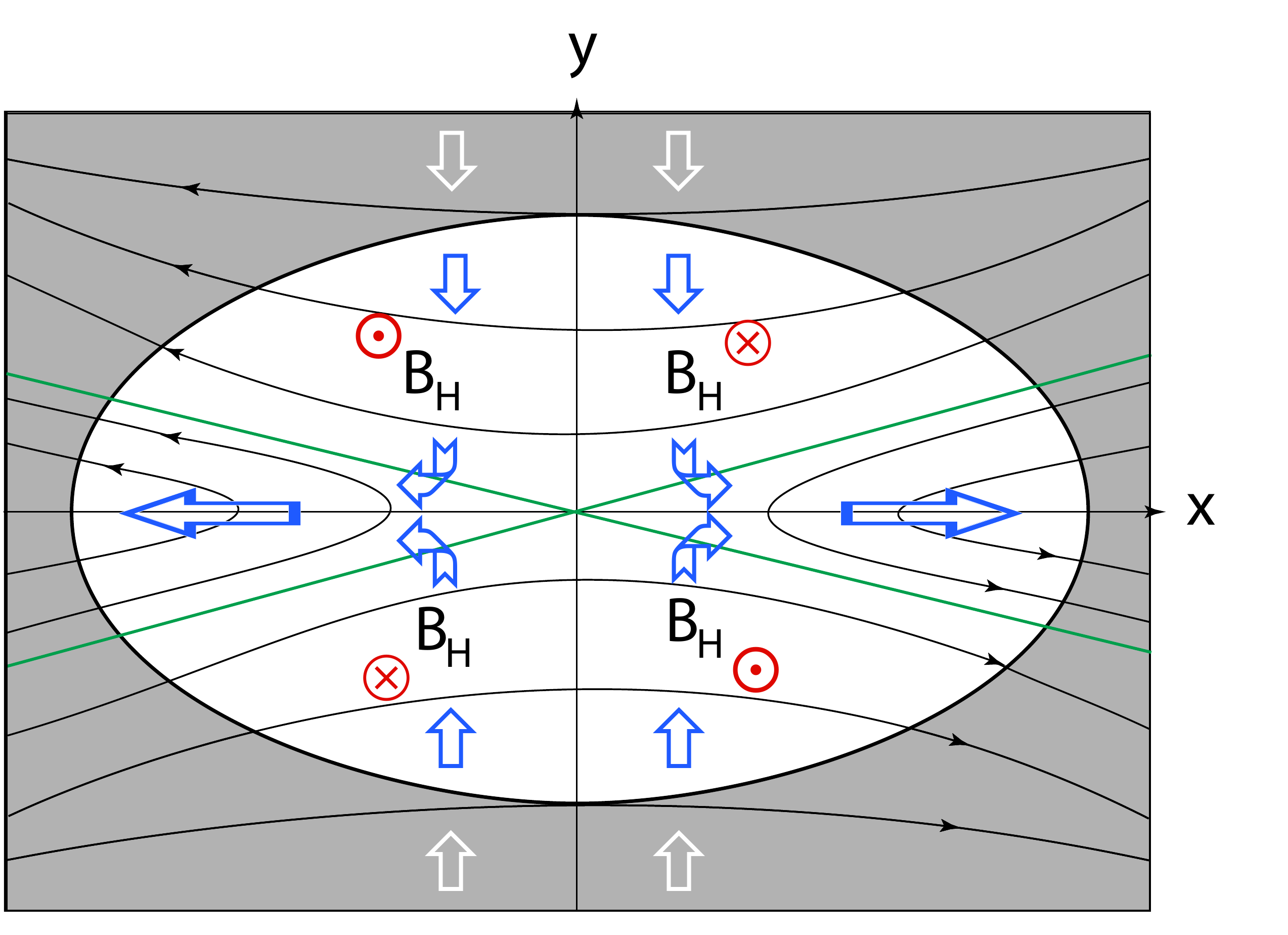}
}}
\caption[ ]
{\footnotesize Schematic geometry of the ion inertial region in the tail current sheet (white domain) around the reconnection {\sf X}-line (seen just as an {\sf X}-point) in collisionless magnetic reconnection. The tail current is not indicated.  In this representation the tail current flows out of the plane while the plasma flows (see the open arrows) in from the top and from the bottom into the centre of the current sheet which includes the {\sf X}-line and, after having crossed the separatrices (green lines),  flows out of the reconnection site (the ion inertial region around the {\sf X}-line) to both sides.  {Note that only a very small fraction of the plasma  flows through the {\sf X}-point; the majority of the plasma that participaties in reconnection flows through the ion inertial region and crosses the separatrices (indicated by bent open arrows). Also shown is the quadrupolar structure of the Hall magnetic field ${\bf B}_H$ indicated by the red circles $\bigodot,\bigotimes$ which have the conventional meaning that the magnetic field vector points either out of or into the plane.}}\label{figrecon-1}
\vspace{-0.3cm}
\end{figure}

Assume that we are dealing with a \emph{stationary} reconnection pattern in the tail current sheet as shown in Figure \ref{figrecon-1} which is part of the tail region in Figure \ref{schematic-fig}. {The reconnection site is centred inside a three-dimensional ion-inertial region (in the figure shown in white) with extension $\lambda_i\lesssim r_{ci}$ in the two directions perpendicular to the magnetic field (where $r_{ci}=v_i/\omega_{ci}$ is the thermal ion gyro-radius, $v_i=\sqrt{T_i/2m_i}$ ion thermal velocity, $T_i$ ion temperature in energy units, $\omega_{ci}=eB/m_i$ ion cyclotron frequency, $m_i$ ion mass) and a distance $\lambda_\|\lesssim v_{i\|}/\omega_{ci}\equiv\beta_{i\|}\lambda_i$ along the magnetic field, where $\beta_{i\|}=2\mu_0NT_{i\|}/B^2$ is the (parallel) ion-plasma-$\beta$ being the ratio of parallel ion thermal $NT_{i\|}$ to magnetic $B^2/2\mu_0$ energy densities. The latter condition takes into account that the ions remain to be unmagnetised along the magnetic field \emph{only} over a distance they can travel with their average parallel thermal speed $v_{i\|}$ within one ion gyro-period.} 

{In Figure \ref{figrecon-1} let the secondary quadrupolar Hall magnetic field (for the global geometry see Figure \ref{schematic-fig}) be ${\bf B}_{\rm H}(x,y)$ which is a function of space and, as noted earlier, is of quadrupolar structure, indicated in the figure by the symbols $\bigodot,\bigotimes$. }Then the Hall-magnetic flux is given by the surface integral of the Hall magnetic field 
\begin{equation}
\Phi_{\rm H}(x,y)=\int\, {\bf B}_{\rm H}\cdot{\rm d}{\bf f}
\end{equation}
which itself is clearly a function of space as well, and ${\rm d}{\bf f}$ is the surface element perpendicular to ${\bf B}_{\rm H}$ (i.e. perpendicular to the plane in Figure \ref{figrecon-1}). The induced electromotive force ${\cal E}_{\rm H}(x,y)$ the Hall magnetic flux $\Phi_{\rm H}(x,y)$ may exert on the plasma is
\begin{equation}
{\cal E}_{\rm H}(x,y)=\int\,{\bf E}_{\rm H}(x,y)\cdot{\rm d}{\bf s} =-\frac{{\rm d}}{{\rm d}t}\Phi_{\rm H}(x,y)
\end{equation}
the line integral of the Hall electric field ${\bf E}_{\rm H}$, which is expressed as the total time derivative of the Hall magnetic flux. Under stationary conditions (non-stationary conditions lead to more complicated pictures and are less transparent; here we assume that the process of reconnection is substantially faster than any typical variation period during a substorm) the plasma convects  at velocity ${\bf V}$ in the frame of the tail current sheet across the Hall region, and the total time derivative reduces to ${\bf V}\cdot\nabla$. {This velocity is to be distinguished from the $\mathbf{E}\times\mathbf{B}$ drift of the electrons which is responsible for the generation of the Hall current. In fact, $\mathbf{V}$ is the velocity difference between the latter and the residual cross-magnetic field inertial velocity of the unmagnetised ions inside the ion inertial region.} One thus has
\begin{equation}\label{eq:emf}
{\cal E}_{\rm H}(x,y)=-{\bf V}\cdot\nabla\Phi_{\rm H}(x,y)
\end{equation}
This electromotive force plays the r$\hat{\rm o}$le of an induced electric  potential that is caused by the mere presence of the Hall magnetic field inside the ion inertial region, {as seen from outside the reconnection site.} 

In the following we show by using a substantially simplified analytical model of both the Hall magnetic field and Hall magnetic flux that the quadrupolar structure of the Hall field just produces the wanted electric potential structure inside the Hall region that maps down to the ionosphere in a way to generate the auroral field aligned electron fluxes.

\subsection*{Simple analytical model}
It requires little sophistication only to see that the Hall magnetic flux itself generates an electromotive force (induced electric potential) of the correct sign for accelerating the ionospheric electron component out of the auroral ionosphere into upward electron fluxes. A very simple model of the Hall magnetic flux suffices for demonstrating this fact. 

Assume that the ion inertial Hall region has rectangular (box) shape in the plane $(x,y)$. In order to approximate the observation that the Hall magnetic field has quadrupolar shape, the Hall magnetic flux can be modelled as
\begin{equation}
\Phi_{\rm H}(x,y)= \Phi_m\sin\left(\frac{\pi [x+\lambda_\|]}{\lambda_\|}\right)\sin\left(\frac{\pi [\lambda_i-y]}{\lambda_i}\right)
\end{equation}
which accounts for an ambient antiparallel magnetic field that is directed in $\mp x$, and $\Phi_m$ is the maximum Hall magnetic flux corresponding to the highest concentration of Hall magnetic field lines pointing either in positive or negative $z$ direction. This  flux is positive (directed out of the plane) in the upper left and lower right quarters of the ion inertial region, it is negative (directed into the plane) in the upper right and lower left quarters (see Figure \ref{figrecon-1}).

For the velocity ${\bf V}$ we assume that in the upper quarters of the box outside the separatrices the flow is directed $-y$, in the lower quarters $+y$, while in the central parts inside the separatrices left of the {\sf X}-point it is directed into $-x$, right of the {\sf X}-point into $+x$. Otherwise the modulus of the velocity is assumed constant. Clearly such a model is oversimplistic, while reproducing the magnetic features of the Hall region. 

Since only gradients parallel to ${\bf V}$ count in the generation of the electromotive force Eq. (\ref{eq:emf}), the derivative jumps from $\nabla_x$ to $\nabla_y$ when passing from the inflow parts to the outflow part of the reconnection site, i.e. when crossing the separatrices. This outflow region is, however, narrow because the plasma is highly accelerated here. We can, therefore, in the simplified approach of our model, safely ignore it in a first discussion before commenting on its presence later on. Moreover, for simplicity we consider only the left upper part of the box located earthward of the {\sf X}-point. Symmetry considerations show that the other quarters behave similarly. 

Performing the differentiation, yields in the inflow region 
\begin{equation}\label{eq:inflow}
{\cal E}_{\rm H}^{\rm in}(x,y)=\frac{\pi V\Phi_m}{\lambda_i}\sin\left(\frac{\pi [x+\lambda_\|]}{\lambda_\|}\right)\cos\left(\frac{\pi [\lambda_i-y]}{\lambda_i}\right)
\end{equation}
In the outflow region one keeps $y$ constant and differentiates with respect to $x$ which, for completeness, yields 
\begin{equation}\label{eq:outflow}
{\cal E}_{\rm H}^{\rm out}(x,y)=\frac{\pi V\Phi_m}{\lambda_\|}\cos\left(\frac{\pi [x+\lambda_\|]}{\lambda_\|}\right)\sin\left(\frac{\pi [\lambda_i-y]}{\lambda_i}\right)
\end{equation}
The numerical factor in front of these expressions determines the real amplitude of the field and is of secondary importance in extracting the physical content of expressions (\ref{eq:inflow}) and (\ref{eq:outflow}).

Below we discuss the obvious implications of this simplified model by applying it to the more realistic elliptical Hall domain. The transfer to another more complicated geometry can be done without any restrictions as only geometric and no physical differences appear in this transfer, which avoids any unjustified mathematical complications. These do not add anything new to the implied physics except for a more precise determination of the boundaries between positive and negative electromotive potentials. Since the model is only approximate and no exact knowledge about the real geometric form of the Hall region is available, more precise mathematics is academic adding only spuriously to the inferences drawn. We intentionally refrain from it in order to avoid any exaggerated (pseudo-)interpretation. 
\begin{figure}[t!]
\centerline{{\includegraphics[width=0.4\textwidth,clip=]{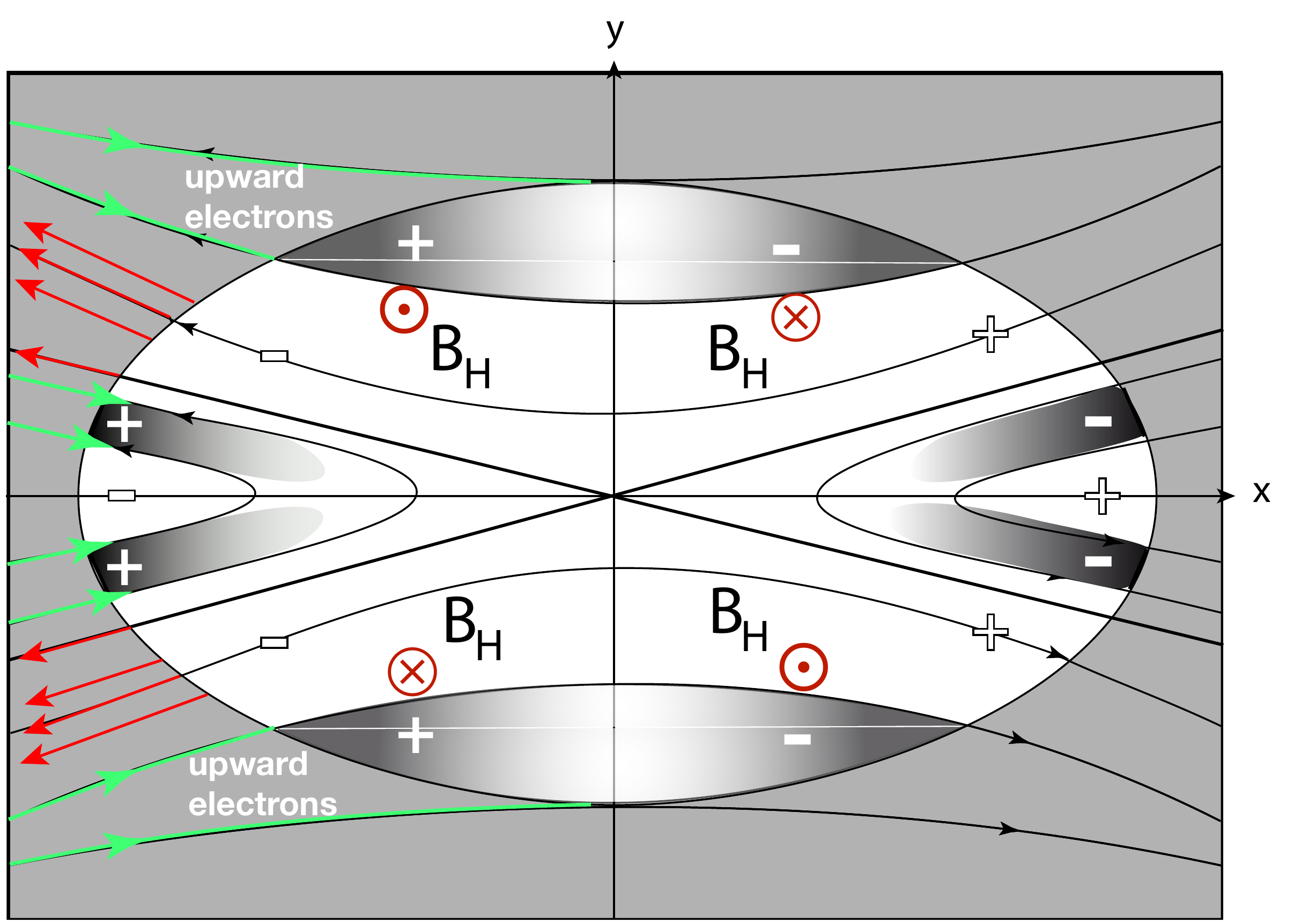}
}}
\caption[ ]
{\footnotesize Schematic of the region of electromotive potentials in the Hall domain of a stationary reconnection region approximately transforming the inference of the simple (rectangular) analytical model to the probable elliptic shape of the reconnection site around the {\sf X}-point. The figure shows the approximate regions of positive electromotive potentials in shading (upward electrons indicated by green arrows). Darker shading means larger potentials. The main domain of positive electromotive potentials is found in the outer part of the inflow region of the Hall domain. Two narrow regions of positive electromotive potential should, however, also exist in the outflow region. The large area left white contains negative electromotive potentials with the exception of a horizontal stripe including the central current sheet where no Hall fields exist. {This stripe is principally empty of any electromotive potentials. Also the very narrow region of maximum Hall fields (not shown) is exempt from any electromotive forces because of the vanishing field gradient. The red arrows indicate down-flowing electrons which have been accelerated during reconnection. The fastes electrons are expected to leave along the separatrices, i.e. from the {\sf X}-line where they have been accelerated out of the current sheet electron reservoir.}}\label{figrecon-2}
\vspace{-0.3cm}
\end{figure}

\subsection*{Discussion and Conclusions}
The inflow-region electromotive potential Eq. (\ref{eq:inflow}) is positive whenever both signs of the trigonometric functions are positive or negative; it will be negative when the signs differ. A positive electromotive potential in the left upper quarter of the inflow region is obtained for $-\lambda_\|<x<0$ and $\lambda_i>y>\frac{1}{2}\lambda_i$ while it becomes negative when $y$ enters the interval $0<y<\frac{1}{2}\lambda_i$. Specular symmetry tells that this behaviour is the same in the entire upper and lower inflow region: Close to the poleward boundary of the ion inertial domain the electromotive force will always be positive.  This is schematically demonstrated in Figure \ref{figrecon-2} where the simple analytical model has formally been transferred to the elliptical shape of the ion-inertial domain. One should, however note, that basing the figure on the symmetric model does not differentiate between earthward and anti-earthward directions in the magnetotail. In the conventional view the Earth is on the left in Figure  \ref{figrecon-2}. Hence only the left-hand part of the figure matters for our purposes. Due to the missing ionosphere and softening of the magnetic field further downtail the right-hand part will favour the evolution of plasmoid-like structures instead.

\begin{figure}[t!]
\centerline{{\includegraphics[width=0.4\textwidth,clip=]{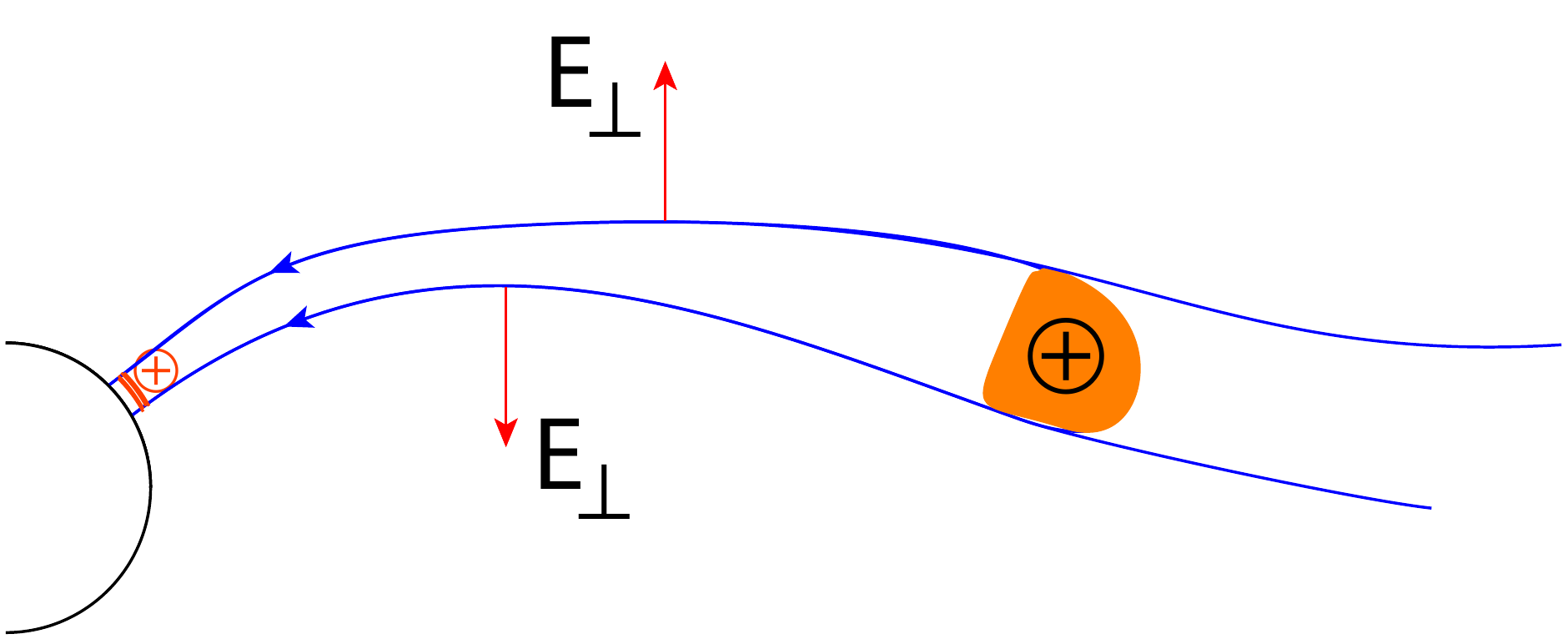}
}}
\caption[ ]
{\footnotesize {A positive electromotive force in the outer magnetosphere enclosed between two field lines is experienced at all altitudes below it as an electric charge. The magnetic field lines behave as equipotentials thus mapping the electromotive difference between them down to the ionosphere where it possesses the same sign but acts like a positive space charge. This implies that the perpendicular electric field along the magnetic flux tube enclosed between the two field lines diverges which causes a shear motion in the surrounding plasma. In this way, the electromotive force in the ion-inertial Hall region  is the source of shear motion along the flux tube and generates the positive topside ionospheric potential and field aligned electric field which extract the ionospheric electron component from the ionosphere and generates a downward field aligned current.}}\label{shear}
\vspace{-0.3cm}
\end{figure}

Magnetic field lines outside the ion-inertial (Hall) region are equipotentials. The shaded zones indicated by  "+" signs are domains of positive electromotive forces, which from the outside cannot be distinguished from positive space charges. { The left-hand side in Figure  \ref{figrecon-2}  maps down to the ionosphere, as shown in Figure \ref{shear},  along the non-reconnected magnetic field. From the positive electromotive potential domain in the inflow region one thus concludes that at the polar boundary of the auroral region a \emph{downward directed electric field} will be seen which may be capable of accelerating the ionospheric electrons upward (green arrows) providing the observed downward auroral currents. Recently, the spatial auroral distribution of these electric fields has bee given \citep{marklund2011}.} 

{We may note that the mapping is a dynamical process which, in reality, is not instantaneous as the stationary picture suggests. Mapping down the potential into the ionosphere proceeds via kinetic Alfv\'en waves. That this is so can be seen from the anticipated scale of the ion inertial region. This scale transverse to the ambient magnetospheric magnetic field is the ion inertial scale $\lambda_i$ which is the transverse scale $k_\perp\lambda_i\sim 1$ of kinetic Alfv\'en waves. The localised electromotive force caused in the Hall region provides the source electric field for kinetic Alfv\'en waves which are launched along the magnetic field and transport the electric distortion caused in the ion inertial region down into the auroral ionosphere.}

The broad inner part of the inflow region (for the moment again ignoring the interruption caused by the presence of the narrow outflow region) is an extended domain of \emph{negative} electromotive force such that the connected ionospheric part sees an \emph{upward} electric field that should cause the ionospheric ions to become accelerated upward, keeping the ionospheric electrons down. This is the upward current region which are constituted by the downward flowing hot magnetospheric electron component part of which comes directly from the reconnection site. This region includes the reconnection site which, however is free of Hall currents and does not give rise to the kind of Hall-induced electromotive potentials. 

Of particular interest is the appearance in the outflow region of narrow domains of positive electromotive potentials located near the separatrix boundary. They result from Eq. (\ref{eq:outflow}) and map down to the lower latitude ionosphere along the reconnected outflow magnetic field. Their presence implies that \emph{the upward current region in an active aurora will always be bounded from both, the polar and the equatorial, sides by comparably narrow regions of upward ionospheric electron fluxes} corresponding to downward current flows. {This is, however, just what is regularly observed in active aurorae during substorms as shown in Figure \ref{aurorec-fig1a}.} 

That our most simple analytical model reproduces this so far unexplained and thus not understood observational fact makes it highly probably that the auroral field-aligned current system is indeed created directly at the reconnection site itself in the near-Earth plasma sheet in the narrow collisionless magnetotail current layer. The vital ingredient of the mechanism that drives these currents is the presence of the Hall-magnetic field in the ion-inertial region. That the Hall field and currents should be involved in the generation of field-aligned currents in the magnetosphere had been conjectured first by \citet{sonnerup1979}. 

We note, finally, the obvious possibility to make use of Eq. (\ref{eq:emf}) for the purpose of an observational determination of the {electromotive force ${\cal E}(x,y)$} in the Hall region with the help of multi-spacecraft missions like Cluster or Themis. For this purpose it suffices to measure the plasma flow velocity and the local Hall magnetic flux.

\vspace{-0.1cm}

\begin{acknowledgements}
RT thanks Andr\'e Balogh, Director at ISSI, for his interest in and dicussions on this subject. 
\end{acknowledgements}

\end{document}